\documentclass{Interspeech2024}
\usepackage{arydshln}

\interspeechcameraready

\title{
Generalized Fake Audio Detection via Deep Stable Learning  
}

\small \name[affiliation={1,2}]{Zhiyong}{Wang}
\small \name[affiliation={1,*}]{Ruibo}{Fu}
\small \name[affiliation={1,}]{Zhengqi}{Wen}
\small \name[affiliation={4}]{Yuankun}{Xie}
\small \name[affiliation={2}]{Yukun}{Liu}
\small \name[affiliation={1,2}]{Xiaopeng}{Wang}
\small \name[affiliation={1}]{Xuefei}{Liu}
\small \name[affiliation={1}]{Yongwei}{Li}
\small \name[affiliation={3}]{Jianhua}{Tao}
\small \name[affiliation={1,2}]{Yi}{Lu}
\small \name[affiliation={1,2}]{Xin}{Qi}
\small \name[affiliation={5}]{Shuchen}{Shi}

\address{
  \small $^1$ Institute of Automation, Chinese Academy of Sciences
  \small $^2$School of Artificial Intelligence, Chinese Academy of Sciences
  \small $^3$Department of Automation and Beijing National Research Center for Information Science and Technology, Tsinghua University \\
  \small $^4$ School of Information and Communication Engineering, Communication University of China 
  \small $^5$ Shanghai Polytechnic University
  \small{\thanks{* corresponding author}}
}
\email{wangzhiyong22@mails.ucas.ac.cn,ruibo.fu@nlpr.ia.ac.cn.}

\keywords{audio spoof, fake audio detection, stable learning, generalization}

\begin{document}

\maketitle

\begin{abstract}
    
Although current fake audio detection approaches have achieved remarkable success on specific datasets, they often fail when evaluated with datasets from different distributions.
Previous studies typically address distribution shift by focusing on using extra data or applying extra loss restrictions during training.
However, these methods either require a substantial amount of data or complicate the training process.
In this work, we propose a stable learning-based training scheme that involves a Sample Weight Learning (SWL) module, addressing distribution shift by decorrelating all selected features via learning weights from training samples.
The proposed portable plug-in-like SWL is easy to apply to multiple base models and generalizes them without using extra data during training.
Experiments conducted on the ASVspoof datasets clearly demonstrate the effectiveness of SWL in generalizing different models across three evaluation datasets from different distributions.


\end{abstract}

\section{Introduction}

Existing synthesis techniques \cite{furuibo01,furuibo02} have the capability to create realistic imitations of human voices.
What is even more concerning is that automatic speaker verification  (ASV) systems \cite{reynolds2002overview} are susceptible to various types of spoofing attacks.
To establish a more reliable automatic speaker verification system, Fake Audio Detection (FAD) model based on machine learning has emerged which aims to discern whether the input audio is produced by synthesis techniques (spoofed) or human (bona-fide). 

While many machine learning approaches perform well under the I.I.D. (i.e., Independent and Identically Distributed) hypothesis, where the evaluation set and training set are independently sampled from the identical distribution. 
For FAD, there will always be new synthesis techniques and unexpected interferences not included in the training set, resulting in an unavoidable distribution shift in the test set.
Therefore, researchers considered the distribution shift scenario when setting up datasets and introduced unseen conditions in the test set.
For instance, ASVspoof2019 \cite{todisco19_interspeech} and ASVspoof2021 \cite{liu2023asvspoof} datasets exhibit significant domain shifts between the training and evaluation sets due to unseen spoofing attack methods and channel variation, making the datasets suitable for training FAD models and evaluating the generalization of FAD models. 

Several studies have explored overcoming the distribution shift conducting experiments on ASVspoof datasets and related literature can be roughly  divided in to two strands. 
One strand explores various training strategies, including multi-task learning \cite{multitasklearning01,multitasklearning02,multitasklearning03}, fusion-based methods \cite{fusion_basedmethods01,fusion_basedmethods02,fusion_basedmethods03}, adversarial learning \cite{adversarial_learning01,adversarial_learning02}, continual learning \cite{continual_learning01,continual_learning02}, transfer learning using large pretrained models \cite{pretrained01,pretrained02,pretrained03}, and contrastive learning \cite{contrastive_learning}. 
This type of method primarily enhances the training constraints by adding additional losses, implicitly supervising the model to obtain more discriminative features. 
While this approach can improve the generalization of the base model, it makes the training process intricate. 
Another strand focuses on the model's input and can be further subdivided into data integration and data augmentation. 
Data integration means co-training a FAD model with multi-dataset, often combined with new training strategies, like \cite{xie23c_interspeech} and \cite{shim23c_interspeech} respectively introduced strategies named aggregation-separation domain generalization method and adaptive sharpness-aware minimization (ASAM). 
The basic idea of data integration is plain and simple, but this approach is sometimes effective and indicates no guarantee of improvements of generalization on every other dataset. Additionally, the interferences caused by the different characteristics of datasets remain unknown.
Data augmentation uses technical means to modify training set data, making the distribution of the training set as close as possible to the distribution of the test set. For example, RawBoost \cite{tak2022rawboost} can be applied to raw audio data to simulate telephone scenarios. 
In \cite{wang2023spoofed}, different vocoders are used to create various spoofing attack types of data based on the training set.
Data augmentation can achieve good results under the premise of knowing the target distribution, which often cannot be satisfied in real applications. 

To address the distribution shift issues and make the FAD model more applicable in more different distributions, this paper proposes a stable learning \cite{Cui2022StableLE} based training scheme that involves a Sample Weight Learning (SWL) module.
The aim of stable learning is to learn a predictive model that can achieve uniformly good performance in any possible environment, which perfectly fits the needs of FAD.
The SWL module can be integrated into existing FAD models. 
This module takes the selected features as input and calculates sample weights to influence the final sample loss. 
Within the SWL module, we employ an approach based on Random Fourier Features (RFF) \cite{RFF} to decorrelate all input features and utilize an iterative optimization approach \cite{iter} to gradually approximate global sample weights by iteratively optimizing local sample weights. 
Importantly, this module can be used as a portable plug-in during training and does not participate in the model's inference process. 
Experimental results demonstrate that the SWL is applicable to multiple base FAD models and achieves consistent performance improvements when evaluated across multiple datasets from different distributions without the need for additional datasets for training or pre-training. 

\begin{figure*}[t!]
  \centering
  \includegraphics[width=\textwidth]{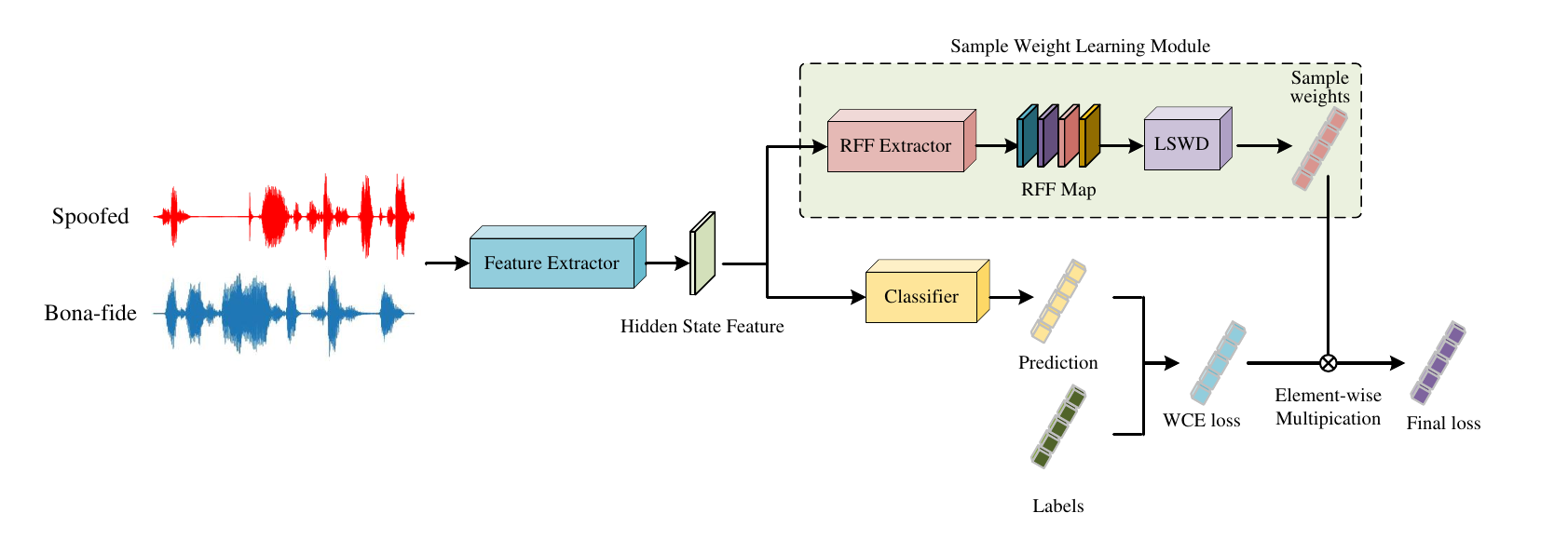}
  \caption[The overall architecture of the proposed stable learning based method.]{The overall architecture of the proposed stable learning based method. LSWD refers to Learning Sample Weighting for Decorrelation as described in Section \protect\ref{sec:sample-weight}. The number of RFF mapping fuctions and hidden state feature are flexible to be adjusted. In the training stage, we only need to feed the selected hidden state feature into SWL module and multiply the computed sample weights with the Weighted Cross-Entropy (WCE) loss.  In the inference phase, the model directly conduct prediction without calculation of sample weights.}
  \label{fig:stable}
\end{figure*}

The contributions of this paper are presented as follows:
\begin{itemize}
\item   We propose a stable-learning based training scheme for FAD called SWL, which involves a SWL module based on RFF and utilizes an iterative optimization strategy. To the best of our knowledge, this is the first study to apply stable learning to improve the generalization of FAD models.
\item Experiments conducted on the ASVspoof datasets show that SWL can generalize AASIST \cite{aasist}, RawNet2 \cite{rawnet2}, and TSSD \cite{tssd} when evaluated across three different datasets.
\item We further explored what combination of node features in the AASIST model might be optimal for decorrelation when applying SWL.  
 \end{itemize}


\section{Proposed Method}

Different from the method mentioned earlier that uses extra data or complicates the training process by computing additional loss, we use stable learning to addressing distribution shift by decorrelating all input features without using extra data.
Figure \ref{fig:stable} shows the overall architecture of proposed SWL.

In stable learning, a sample weighting method \cite{kuang2020stable} is proven to be useful to decorrelate features and help linear models produce more stable predictions under distribution shift.
When extending these ideas into FAD models to handle more complex data types like audio, we first need to consider how to quantify independence between features. 
Moreover, this sample weighting method requires meticulous calculations for all features, which is not suitable for training deep neural networks. 

In the following subsections, we detail sample weighting with RFF and iteratively learn global sample weights, which correspond to solving the two problems mentioned above. 

\subsection{Sample weighting with RFF}
\label{sec:sample-weight}
In stead of focusing only on eliminating dependence between bona-fide and spoofed samples, we tend to eliminate dependence between all input features so that we can get more dispersed representations of all samples.
To eliminate the dependence between any pair of features $Z_{:,i}$ and  $Z_{:,j}$ from samples in the representation space, we first need to quantify the independence  between features.
In \cite{bahng2020learning}, Hilbert-Schmidt Independence Criterion (HSIC) is proven to be capable of applying as a criterion to supervise feature decorrelation, but HSIC computationally cost too much on large datasets.  In Euclidean space, Frobenius norm corresponds to the HilbertSchmidt norm \cite{strobl2019approximate}, in this paper,  Frobenius norm is used to quantify the independence between features.

We sample $(A_1, A_2, \ldots, A_n)$ and $(B_1, B_2, \ldots, B_n)$ from the distribution of $A$ and $B$. 
Consider a measurable, positive definite kernel $k_D$ on the domain of random variable $D$ and we denote the corresponding Reproducing Kernel Hilbert Space as $\mathcal{H}_D$. 
If we denote cross-covariance operator from $\mathcal{H}_B$ to $\mathcal{H}_A$ by the symbol $\Sigma_{\text{AB}}$, then the partial cross-covariance matrix will be:
 
\begin{align}
\label{equation:eq1}
\hat{\Sigma}_{AB} &= \frac{1}{n-1} \sum_{i=1}^{n} [
(\textbf{u}(A_i)- \frac{1}{n} \sum_{j=1}^{n} \textbf{u}(A_j))^T \cdot \nonumber \\
&\phantom{=AAAA==}  (\textbf{v}(B_i)- \frac{1}{n} \sum_{j=1}^{n} \textbf{v}(B_j)) 
],
\end{align}
where
\begin{align}
\textbf{u}(A) &= (u_1(A), u_2(A), \ldots, u_{n_A}(A)),\quad u_j(A) \in \mathcal{H}_\text{RFF},\quad \forall j, \nonumber \\
\textbf{v}(B) &= (v_1(B), v_2(B), \ldots, v_{n_B}(B)),\quad v_j(B) \in \mathcal{H}_\text{RFF},\quad \forall j,
\end{align}

Here $\textbf{u}$ and $\textbf{v}$ are the RFF mapping functions, $n_A$ and $n_B$ represent the numbers of functions from $\mathcal{H}_\text{RFF}$ and $\mathcal{H}_\text{RFF}$ denotes the function space of RFF with the following form:

\begin{align}
\mathcal{H}_\text{RFF} &= \{ h : x \rightarrow \sqrt{2} \cos(\omega x + \phi) \mid \nonumber \\
&\phantom{==} \omega \sim {N}(0, 1), \phi \sim \text{Uniform}(0, 2\pi) \},
\end{align}
where $\omega$ and  $\phi$ are sampled from the standard Normal distribution and the Uniform distribution respectively.
Then, the independence quantification metric $I_{AB}$ is defined as the Frobenius norm of the partial cross-covariance matrix ($I_{AB} = \Vert \hat{\Sigma}_{{AB}} \Vert_F^2$).
Note that $I_{AB}$ is always non-negative. So the lower the $I_{AB}$, the more independent the two variables $A$ and $B$.

After establishing the quantification criteria, we can then measure general independence via RFF and eliminate the dependence between hidden state features through sample weighting.
We denote the sample weights as $w \in \mathbb{R}_+^n$ and $\Sigma_\text{i=1}^n \omega_i=n$.
After weighting, the $ \hat{\Sigma}_{\text{AB}}$ changes into $ \hat{\Sigma}_{\text{AB;w}}$  which can be calculated using Equation \eqref{equation:eq1} as follows:
 \begin{align}
\label{equation:abw}
\hat{\Sigma}_{AB;\text{w}} &= \frac{1}{n-1} \sum_{i=1}^{n} [
( \omega_i \textbf{u}(A_i)- \frac{1}{n} \sum_{j=1}^{n}  \omega_j 
 \textbf{u}(A_j))^T \cdot \nonumber \\
&\phantom{=AAAA==}  ( \omega_i \textbf{v}(B_i)- \frac{1}{n} \sum_{j=1}^{n}  \omega_j  \textbf{v}(B_j)) 
],
\end{align}

Specifically, for selected features $Z_{\text{:,i}}$ and  $Z_{\text{:,j}}$ mentioned above, we need to optimize $\textbf{w}$ by
\begin{align}
  \textbf{w}^* &= \underset{\textbf{w}\in \Delta_n}{arg min} \underset{1\leq i < j \leq m_z}{\sum}\Vert \hat{\Sigma}_{{Z_{:,i},Z_{:,j};\textbf{w}}} \Vert_F^2,
\end{align}
where $\Delta_n = \{\textbf{w} \in \mathbb{R}_+^n \mid \sum_{i=1}^n w_i=n\}$. Hence, weighting training samples with the optimal $\textbf{w}^*$ can mitigate the dependence between all selected hidden state features.

Generally, this algorithm iteratively optimize sample weights $\textbf{w}$, feature extractor function $f$ , and classifier function $g$ as follows:
\begin{align}
\label{equation:train}
  f^{(t+1)},g^{(t+1)} &= \underset{f,g}{arg min} \sum_{i=1}^{n} w_i^{(t)} L(g(f(X_i)),y_i), \nonumber \\ 
  \textbf{w}^{(t+1)} &= \underset{\textbf{w}\in \Delta_n}{arg min} \underset{1\leq i < j \leq m_z}{\sum}\Vert \hat{\Sigma}_{{Z^{(t+1)}_{:,i},Z^{(t+1)}_{:,j};\textbf{w}}} \Vert_F^2,
\end{align}
 where $Z^{(t+1)} = f ^{(t+1)}(X)$,  $X_i$ and $y_i$ means an input audio and prediction of the classifier, $ L(·, ·)$ represents the weighted cross entropy loss function and  $t$ represents the time stamp. Initially, $\textbf{w}^{(0)} = (1, 1, . . . , 1)^T$ .
 
\subsection{Iteratively learn global sample weights}
 
Equation \ref{equation:train} requires weight learned for every single  input data. However,  it cost too much for a FAD model based on deep neural networks trained on a large dataset to learn sample weights globally.  For every batch during training, the features and the sample weights  appear as follows:  
\begin{align}
\text{Z}_O, &= \text{Concat}(\text{Z}_{G1},\text{Z}_{G2}, \ldots,\text{Z}_{Gk},\text{Z}_{L}) \nonumber \\
\text{w}_O, &= \text{Concat}(\text{w}_{G1},\text{w}_{G2}, \ldots,\text{w}_{Gk},\text{w}_{L}) 
\end{align}
$\text{Z}_O$ and $\text{w}_O$, represent the current features and weights, are used to optimize the new sample weights in the next batch.   
$\text{Z}_{G1}, \ldots,\text{Z}_{Gk}$ and $\text{w}_{G1}, \ldots,\text{w}_{Gk}$ means features and weights represent the former 
global information learned from all previous batches.  
 $\text{Z}_L$ and $\text{w}_L$  means the features and weights in the current batch, representing local information. When a FAD model is trained with a large dataset, the concatenation would incur significant space consumption and lead to an increase in computational complexity. 
 
 To address this problem, we utilize an approximation method proposed in \cite{iter}  which fuses the former global information and the local information at the end of each epoch as follows:
\begin{align}
\text{Z}_{G_i}^{'}, &= \alpha_i\text{Z}_{Gi} + (1-\alpha_i)\text{Z}_L  \nonumber \\
\text{w}_{G_i}^{'}, &= \alpha_i\text{w}_{Gi} + (1-\alpha_i)\text{w}_L 
\end{align}
We substitute $(\text{Z}_{Gi}, \text{w}_{Gi})$ with $(\text{Z}_{Gi}^{'}, \text{w}^{'}_{Gi})$ for the next batch. and use the hyperparameter $\alpha$ to control the information, where a large value indicates long-term memory in the information, and a smaller value indicates short-term memory. In this paper, we set $\alpha$ to $0.9$.
Detailed procedure can be found in Appendix A.1 of \cite{iter}, we use Adam optimizer instead of SGD for optimization to converge faster and more stable.

\section{Experiments}
\subsection{Datasets and Evaluation metrics}
All experiments are trained on the Logical Access (LA) subset of the ASVspoof 2019 dataset. 
To verify the generalization of models, we utilized the ASVspoof 2021 dataset, which only contains evaluation sets and is the latest and most challenging edition of the ASVspoof challenge series.
In comparison to ASVspoof 2019, the utterances in the LA scenario are transmitted across real telephone systems and the utterances in the DF scenario are processed through various audio compressors.  Table~\ref{tab:dataset} presents the overall statistics of the datasets used in the experiments.

Equal error rate (EER) is used to evaluate the performance of models. The lower the EER value, the better the models.
\begin{table}[h]
  \caption{Datasets used in the experiments. \# Spks, \# Utts, and \# Attks indicate the number of speakers, utterances, and spoofing attacks, respectively. Training and evaluation set is divided by {/}. 19LA, 21LA and 21DF respectively represent ASVspoof 2019 LA, ASVspoof 2021 LA, ASVspoof 2021 DF.}
  \label{tab:dataset}
  \centering
  \begin{tabular}{l c c c}
    \toprule
    \textbf{Dataset}      & \textbf{\# Spks}     & \textbf{\# Utts}     & \textbf{\# Attks}     \\
    \midrule
    19LA              & 20 / 48  & 25380 / 108978       &6 / 13                  \\
    21LA               & - / 48  &- / 181566      &- / 13                  \\
    21DF       & - / 48  & - / 611829      &- / 13               \\
    \bottomrule
  \end{tabular}
\end{table}
\subsection{Experimental setup}
For experiments, we choose three FAD models as base models to apply SWL, and we will briefly explain why we choose them and how we apply SWL to each of them.

First base FAD model is AASIST model, both the original version and the light version  respectively denoted as AASIST and AASIST-L. 
There two reasons for choosing this model. 
Firstly, this model is open source and has been recently utilized in multiple research studies, demonstrating state-of-the-art (SOTA) performance.
Secondly, the features sent into the classifier in this model are composed of multiple node features extracted based on temporal and spectral domains. This setup allows for further exploration of feature interpretability regarding the optimal combination of node features for decorrelation.

Rawnet2 is well-known as one of the baselines of the ASVspoof challenge and a typical end to end model.
As for the TSSD model, its performance is SOTA among end-to-end models.
We use the input of the last fully connection layer in both models to apply SWL.
We do not adjust parameters (e.g. learning rate, model architecture configuration) of all three base FAD models. 
The best models are chosen based on the lowest EER on the development set of 19LA.

\begin{table}[t!]
  \caption{EER(\%) results of the base FAD models and applying SWL to them. The results about ASAM are from \cite{shim23c_interspeech} and the $^*$ represents the FAD model is trained on multi-dataset.}
  \label{tab:improvebasemodel}
  \centering
  \begin{tabular}{l c c c}
    \toprule
    \textbf{Model}      & \textbf{19LA}     & \textbf{21LA}     & \textbf{21DF}     \\
    \midrule
    AASIST              & 1.52  &9.96       &20.45                  \\
    AASIST + SWL     & \textbf{1.38}  &\textbf{9.40}       &\textbf{19.09}                 \\
    \midrule
    AASIST-L            & 1.26  &11.44      &22.40                 \\
    AASIST-L + SWL   & \textbf{1.14}     &\textbf{8.02}       &19.67      \\
    \hdashline
    AASIST-L + ASAM       &1.48   &10.18      &\textbf{19.58}         \\
    AASIST-L + \text{ASAM$^*$}  &1.27   &12.41      &19.84               \\
    \midrule
    RawNet2             & 4.72  &11.04      &22.13                  \\
    RawNet2 + SWL    & \textbf{4.47}  &\textbf{8.69}       &\textbf{21.14}                 \\
    \midrule
    TSSD                & 2.12  &17.75      &32.10                  \\
    TSSD + SWL       & \textbf{2.00}  &\textbf{15.83}      &\textbf{29.55}               \\
    \bottomrule
  \end{tabular}
\end{table}

\section{Results and Analyses}

\subsection{SWL generalizes base FAD models} 
In Table\ref{tab:improvebasemodel}, we validate the proposed SWL method by evaluating models on three datasets from different distributions, namely 19LA, 21LA and 21DF. 
From the results, it can be observed that WSL is able to simultaneously generalize the base FAD models on multiple datasets from different distributions.  

\begin{figure}[b!]
  \centering
  \includegraphics[width=\linewidth]{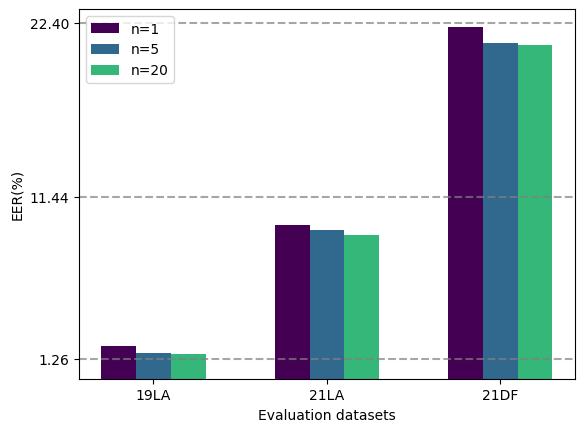}
  \caption{Applying SWL to AASIST-L using different numbers of RFF mapping functions. The dashed line represents the performance of AASIST-L without applying SWL. }
  \label{fig:numberofRFFf}
\end{figure}
In addition, We compare SWL with another method known as ASAM, which has shown significant advancements in enhancing the generalization of base FAD models \cite{shim23c_interspeech}.
The results demonstrate that SWL achieves competitive performance comparing ASAM in assisting the base model when trained on the same single dataset.
Furthermore, upon comparing the performance of the base model using ASAM and training it with multiple datasets, it becomes evident that SWL can better generalize the base model without the need for extra training data.

\subsection{More RFF mapping functions, higher generalization}
In Section \ref{sec:sample-weight}, we demonstrate how proposed SWL method work theoretically. 
There is a hyperparameter that might influence the performance, which is the number of RFF mapping functions. 
In this part, we conduct some experiments applying SWL to AASIST-L model selecting all node features to be decorrelated and using different numbers of RFF mapping functions.
The results are recorded in the Figure \ref{fig:numberofRFFf}.

It turns out that, for AASIST-L model, generalization is better when number of RFF mapping functions is higher, but at the same time, both the training duration and computational complexity will also increase. 
In additional, there is a decrease in performance of all models on 19LA.
We attribute this to the non-optimal combination of node features selected for decorrelation, and we will discuss it in the next section.

\begin{table}[t]
  \caption{Different combination of node features applying SWL to AASIST-L model. }
  \label{tab:comb}
  \centering
  \begin{tabular}{l c c c}
    \toprule
    \textbf{Node Features}& \textbf{19LA}     & \textbf{21LA}     & \textbf{21DF}     \\
    \midrule
    5N&1.55       &9.02       &20.99  \\
    4N     &1.15       &11.68      &20.24  \\
    2N-T     &1.40       &10.23       &21.22  \\
    2N-S     &\textbf{1.14}       &\textbf{8.02}       &\textbf{19.67}  \\
    \bottomrule
  \end{tabular}
\end{table}
\subsection{What combination of nodes is better for decorrelation}
In AASIST framework \cite{aasist}, the temporal and spectral domains are modeled by two graph moduls in parallel and finally turn to four node features representing temporal and spectral information. Finally, the four node features and a stack node features are sent to classifier after simple concatenation. 

In this section, we explore what combination of the five node features is optimal to be selected for feature decorrelation. 
Four combinations were considered, namely, all five node features (5N), removing the stack node features (4N), two node features related to spectral information (2N-S), and two node features related to temporal information (2N-T). 
Then, we conduct experiments based on AASIST-L model setting the number of RFF mapping functions to 20. The experimental results in Table \ref{tab:comb} show that, among the four combinations, 2N-S is optimal for feature decorrelation and yields the best results, indicating that decorrelating the 2N-S is easier, and the spectral domain is more helpful for  FAD. 


\section{Conclusions}
In this paper, we propose a stable learning based training scheme named SWL for FAD, which involves a sample weight learning module based on Random Fourier Features and utilizes an iterative optimization strategy. 
SWL focuses more on selected feature that extracted by FAD models and decorrelates them to help model produce more stable predictions under distribution shift.
Experiments conducted on the ASVspoof datasets demonstrate that our methods can generalize multiple FAD models when evaluated across three different datasets.
We further find out that, for AASIST-L, generalization becomes better when the number of RFF mapping functions is higher, and the node features related to spectral features are better for decorrelating when applying SWL. 

\section{Acknowledgements}
This work is supported by the National Natural Science Foundation of China (NSFC) (No.62101553, No.62306316, No.U21B20210, No. 62201571).

\tiny
\bibliographystyle{IEEEtran}
\bibliography{main}

\end{document}